\documentclass[a4paper,10pt]{article}
\usepackage{graphicx}
\usepackage{amsmath}

\usepackage[sort&compress,numbers]{natbib}
\usepackage{amsfonts}
\usepackage{amssymb}
\usepackage{amsmath}
\usepackage{latexsym}
\usepackage{color}
\usepackage{ntheorem}
\usepackage{braket}
\usepackage[colorlinks=true]{hyperref}
\usepackage{authblk}

\newcounter{mnotecount}[section]

\renewcommand{\themnotecount}{\thesection.\arabic{mnotecount}}
\newcommand{\mnote}[1]
{\protect{\stepcounter{mnotecount}}$^{\mbox{\footnotesize
$
\bullet$\themnotecount}}$ \marginpar{
\raggedright\tiny\em
$\!\!\!\!\!\!\,\bullet$\themnotecount: #1} }

\newcommand{\tim}[1]{\mnote{{\bf tim:}#1}}

\newcommand{\jlcax}[1]{}
\newcommand{\eean}{\nonumber\end{eqnarray}}

\newcommand{\kk}[1]{}

\newcommand{\beq}{\begin{equation}}

\newcommand{\FS}       
                  {F}

\newcommand{\HS} 
       {H_{\mbox{\scriptsize volume}}}

{\ptc{this should be removed in the oberwolfach version}}%

\newcommand{\eeal}[1]{\label{#1}\end{eqnarray}}
\newcommand{\bed}{\begin{deqarr}}
\newcommand{\eed}{\end{deqarr}}
\newcommand{\bedl}[1]{\begin{deqarr}\label{#1}}
\newcommand{\eedl}[2]{\arrlabel{#1}\label{#2}\end{deqarr}}

\newcommand{\bel}[1]{\begin{equation}\label{#1}}
\newcommand{\bea}{\begin{eqnarray}}
\newcommand{\bean}{\begin{eqnarray}\nonumber}
\newcommand{\beal}[1]{\begin{eqnarray}\label{#1}}
\newcommand{\eea}{\end{eqnarray}}

\newcommand{\be}{\begin{equation}}
\newcommand{\eeq}{\end{equation}}
\newcommand{\ee}{\end{equation}}
\newcommand{\beqa}{\begin{eqnarray}}
\newcommand{\eeqa}{\end{eqnarray}}
\newcommand{\beqan}{\begin{eqnarray*}}
\newcommand{\eeqan}{\end{eqnarray*}}
\newcommand{\ba}{\begin{array}}
\newcommand{\ea}{\end{array}}

\newcommand{\scri}{{\mycal I}}%

\newcommand{\warn}[1]
{\protect{\stepcounter{mnotecount}}$^{\mbox{\footnotesize
$
\bullet$\themnotecount}}$ \marginpar{
\raggedright\tiny\em
$\!\!\!\!\!\!\,\bullet$\themnotecount: {\bf Warning:} #1} }

\newcommand{\eq}[1]{(\ref{#1})}

\newcommand{\ptc}[1]{\mnote{{\bf ptc:}#1}}

\newcommand{\beqar}{\begin{deqarr}}
\newcommand{\eeqar}{\end{deqarr}}

\newcommand{\beaa}{\begin{eqnarray*}}
\newcommand{\eeaa}{\end{eqnarray*}}

\def\H{\mathcal{H}}

\DeclareFontFamily{OT1}{rsfs}{}
\DeclareFontShape{OT1}{rsfs}{m}{n}{ <-7> rsfs5 <7-10> rsfs7 <10-> rsfs10}{}
\DeclareMathAlphabet{\mycal}{OT1}{rsfs}{m}{n}

{\catcode `\@=11 \global\let\AddToReset=\@addtoreset}
\AddToReset{equation}{section}

{\catcode `\@=11 \global\let\AddToReset=\@addtoreset}
\AddToReset{figure}{section}

{\catcode `\@=11 \global\let\AddToReset=\@addtoreset}
\AddToReset{table}{section}

\begin{document}

\title{The limit of Kerr-de Sitter spacetime with infinite angular-momentum parameter $a$}
\author[1]{Marc Mars}
\author[2]{Tim-Torben Paetz}
\author[3]{Jos\'e M. M. Senovilla}
\affil[1]{Instituto de F\'isica Fundamental y Matem\'aticas, Universidad de Salamanca, Plaza de la Merced s/n, 37008 Salamanca, Spain}
\affil[2]{Gravitational Physics, University of Vienna, Boltzmanngasse 5, 1090 Vienna, Austria}
\affil[3]{F\'isica Te\'orica, Universidad del Pa\'is Vasco, Apartado 644, 48080 Bilbao, Spain}

\maketitle

\vspace{-0.2em}

\begin{abstract}
We consider the limit $a\rightarrow \infty$ of the Kerr-de Sitter spacetime. The spacetime is a Petrov type-D solution of the vacuum Einstein field equations with a positive cosmological constant $\Lambda$, vanishing Mars-Simon tensor and conformally flat $\scri$. It possesses an Abelian 2-dimensional group of symmetries whose orbits are spacelike or timelike in different regions, and it includes, as a particular case, de Sitter spacetime. The global structure of the solution is analyzed in detail, with particular attention to its Killing horizons: they are foliated by non-compact marginally trapped surfaces of finite area, and one of them `touches' the curvature singularity, which resembles a null 2-dimensional surface. Outside the region between these horizons there exist trapped surfaces that again are non-compact. The solution contains, apart from $\Lambda$, a unique free parameter which can be related to the angular momentum of the non-singular horizon in a precise way. A maximal extension of the (axis of the) spacetime is explicitly built. We also analyze the structure of $\scri$, whose topology is $\mathbb{R}^3$.
\end{abstract}


\tableofcontents



\section{Introduction}
In \cite{mpss} vacuum spacetimes with a positive cosmological constant admitting a conformal compactification and a Killing vector field whose associated Mars-Simon tensor vanishes were studied. Such spacetimes were called {\em Kerr-de Sitter like} as they contain the family of Kerr-NUT-de Sitter spacetimes and related metrics \cite{mars_senovilla}. The conformal compactification implies the existence of null infinity $\scri$, which is a spacelike hypersurface in the conformally extended spacetime, and the Killing vector extends smoothly as a tangential conformal Killing vector field at $\scri$. In \cite{mpss} a very specific and explicit structure of the Cotton-York tensor of $\scri$ was found in terms of that conformal Killing vector.

Then, the complete classification of Kerr-de Sitter like spacetimes with conformally flat $\scri$ (that is, with vanishing Cotton-York tensor) was carried out in \cite{mpss2}. This classification led us, somewhat unexpectedly, to the following line element,
 \begin{equation}
g \,=\, \Delta^{\infty}\Big(\mathrm{d}t -  \sin^2\theta\mathrm{d}\phi\Big)^2 + \Sigma^{\infty}\sin^2\theta\Big( \mathrm{d}t-( r^2+1) \mathrm{d}\phi\Big)^2
- \frac{\mathrm{d}r^2}{\Delta^{\infty}}+\frac{\mathrm{d}\theta^2}{\Sigma^{\infty}}
\;,
\label{KdS_limit}
\end{equation}
with
\begin{eqnarray*}
\Sigma^{\infty}&:=& \frac{\Lambda \cos^2\theta}{ 3(r^2 +\cos^2\theta)}
\;,
\\
\Delta^{\infty} &:=&\frac{\Lambda r^2(1+ r^2)+ 6 m r}{3(r^2 +\cos^2\theta)}
\;,
\end{eqnarray*}
which is a solution of the vacuum Einstein equations with a positive cosmological constant $\Lambda$ 
\begin{equation}
R_{\mu\nu}=\Lambda g_{\mu\nu}\,, \quad \Lambda>0 
\, , \label{EFE}
\end{equation}
and where $m$ is a free parameter.
 It must be observed that the cosmological constant cannot vanish, in other words, there is no analogue of this spacetime starting from the Kerr 
metric. No analogue with negative cosmological constant exists 
either. The parameter
$\Lambda$ in the metric \eq{KdS_limit} can be taken to be negative, but then
the signature of the metric changes and a global sign must be introduced
to restore the original signature.
Given that the Ricci tensor remains unaffected by this sign, equation
\eq{EFE} implies that the newly constructed metric is still an
Einstein space with positive cosmological constant. In fact, it is isometric
to the starting metric  with the opposite value of $m$. 

In \cite{mpss2} we called this solution the  \emph{$a\rightarrow \infty$-KdS-limit-spacetime} because it can be obtained as a limit of the Kerr-de Sitter (KdS) family when the parameter $a$ ---the angular momentum per unit mass--- goes to infinity. See Section \ref{extension} for a similar derivation using the same limit.

The line-element \eq{KdS_limit} is a member of the Plebanski-Demianski family of Petrov type D spacetimes ---and as such this metric was already `known'---, though, to our knowledge, it has not been discussed in the literature so far. Due to its relationship to the Kerr-de Sitter family, and to some rather unusual properties that we will encounter later, it seems to have some physical and theoretical interest, as well. It is the aim of this note to provide a discussion of some of its properties.

We start with the basic features of the solution in Section \ref{Basics}. In particular, we identify a curvature singularity, the axis of symmetry, and the horizons through which the metric can be extended. After providing an extension of (\ref{KdS_limit}) by performing the limit $a\rightarrow \infty$ to the Kerr-de Sitter metric written in Kerr coordinates in Section \ref{extension}, we also present an alternative expression of the metric in generalized Kerr-Schild form with de Sitter spacetime as seed metric (subsection \ref{Kerr-Schild}). In section \ref{TrappedSurfaces} we show that the mentioned horizons are actually Killing horizons and identify the causal character of the orbits of the 2-dimensional group of motions and the existence of (marginally) trapped surfaces. We also find that the axial Killing vector becomes timelike in some regions leading to the existence of closed timelike curves.  

Section \ref{KillingHorizons} is devoted to a deeper analysis of the Killing horizons, we find their surface gravity, study the geometry of their marginally trapped cuts, and we also compute the angular momentum of the horizon not touching the singularity. A brief discussion about the possible mass of the solution is included. In section \ref{extensions2} a maximal extension of the spacetime is performed along the axis of symmetry. Finally section \ref{Scri} is devoted to the intrinsic properties of infinity, $\scri$, and its topology.

\section{$a\rightarrow \infty$-KdS-limit-spacetime: basic features}
\label{Basics}
In this section we present the main features of the spacetime, fix the coordinate ranges and the physical units of the parameter $m$ and of the coordinates. 
To start with, we note that the metric (\ref{KdS_limit}) has two obvious Killing vector fields, $\partial_\phi$ and $\partial_t$, and they commute. It follows from the considerations in \cite{mpss2} that there are no further Killing vectors whenever $m\neq 0$. The case $m=0$ leads to a maximally symmetric spacetime and therefore it is (possibly a portion of) de Sitter spacetime. 
We also note that the metric has a symmetry 
\be
(m,r) \leftrightarrow (-m,-r) . \label{symmetry}
\ee


For later use we determine the physical dimensions of the various quantities
in \eq{KdS_limit}. Since the line-element $g$ has dimensions of square length,
$L^2$, and $[\Lambda] = L^{-2}$ it follows immediately that $[t] = L^2$,
$[\phi] = L^2$, $r, \theta$ are adimensional and 
$[m] = L^{-2}$.

In order to fix the allowable range of the coordinates $\{r,t,\theta,\phi\}$ we are using, we first of all observe that there are obvious problems at:
\begin{itemize}
\item $\theta =0$,
\item $\theta=\pi/2 \Longrightarrow \Sigma^\infty =0$,
\item $r^2+\cos^2\theta =0$,
\item $\Delta^\infty =0 \Longrightarrow \Lambda r^2(1+ r^2)+ 6 m r =0$.
\end{itemize}
We consider these in turn.

The $\theta=0$ problem indicates the existence of an axis of symmetry \cite{MS}. The Killing vector $\partial_\phi$ vanishes there ---and there exist regular coordinates where this can be easily proven---, and the regularity condition on the axis \cite{MS} is satisfied by simply choosing as axial Killing vector 
\begin{align*}
\eta := \frac{3}{\Lambda} \partial_{\phi}.
\end{align*}
Then, the $2\pi$-periodicity for the angular coordinate $(\Lambda/3)\phi$ is well defined, and this Killing vector has closed orbits. Observe that, as usual, even though $\partial_\phi$ is spacelike (as it must) nearby the axis $\theta \rightarrow 0$, these closed orbits may lead to the existence of closed timelike curves elsewhere. This will actually happen, see below.

The case of $\theta =\pi/2$ is of a different kind. One can check that, due to \eq{EFE}, the only relevant part of the curvature is given by the Weyl tensor, but as the Petrov type is D it can be proven that the only non-vanishing Weyl scalar is
$$
\Psi_2 = \frac{m}{(r+i \cos\theta )^3}
$$
so that the only relevant curvature invariants are given by
\begin{equation}
I =3\Psi_2^2= \frac{3 m^2}{(r+ i \cos\theta)^6}
\,. \label{invariant}
\end{equation}
This implies that $\cos\theta =0$ has no curvature problem (unless $r=0$). Actually, by inspection of the metric, one notices that $\cos\theta=0$ seems to sit at infinite distance, because a line with constant values of $r\neq0$, $t$ and $\phi$ has an infinite length to reach $\theta =\pi/2$. This suggests that a preferable coordinate is given by
\be
dx := \frac{1}{\cos\theta}d\theta\hspace{3mm} \Longrightarrow \hspace{3mm}  \sin\theta =\tanh x \label{x-coord}
\ee
so that the axis corresponds to $x\rightarrow 0$ while $\theta =\pi/2$, which is at infinity for $r\neq 0$, corresponds to $x\rightarrow \infty$.

Keeping $\theta =\pi/2$, there remains the case with $r=0$ too, which is the third bullet above. From \eq{invariant} it follows that this is a curvature singularity ---unless, of course, $m=0$, the de Sitter case. It is important to remark that now $\theta =\pi/2$ is no longer at infinity, because $\Sigma^\infty (r=0) =\Lambda/3$. Therefore, the region with $r=\cos\theta =0$, that is
\begin{equation}
\{r=0,\,  \theta= \pi/2\}
\,.\label{singularity}
\end{equation}
must be cut out from our manifold. Actually, the entire zone with $r=0$ is not covered by the present coordinates, as we are going to see presently, and thus this will become relevant in the next section, after we perform appropriate extensions. 

The fourth and final bullet above hints at the definition of the following function of $r$
\be
F(r):= \frac{\Lambda}{3}r^2(r^2+1)+2mr \label{F}
\ee
which will be used later on. With this definition we have
$$
\Delta^\infty = \frac{F(r)}{r^2+\cos^2\theta}
$$
and, due the form \eq{KdS_limit}, we must choose either the region with $F>0$ or with $F<0$. In the former case $r$ is a time coordinate, while in the second is a spatial one. The function $F(r)$ has the following properties (primes denote derivative with respect to $r$):
$$
F(r\rightarrow \pm \infty) = + \infty, \hspace{4mm} F(0)=0,\hspace{4mm} F''(r)=4\Lambda r^2 +\frac{2\Lambda}{3} >0
$$
from where it easily follows that there is a unique value $\bar r$ such that $F'(\bar r)=0$, where $F(r)$ has the unique and absolute minimum and $F(\bar r)<0$ (unless $m=0$). Hence, there exists precisely one value $r_1 \neq 0$ satisfying $F(r_1)=0$. Taking into account that $F'(0)=2m$, the sign of $m$ decides the signs of $\bar r$ and of $r_1$. If, say, $m>0$ then both $\bar r$ and $r_1$ are negative. Due to the symmetry \eq{symmetry} the opposite happens if $m<0$. 

Given the definition of $r_1$ we can replace, if needed, the constant $m$ by $r_1$ everywhere by noting (we assume momentarily $r_1 \neq 0$ i.e. $m\neq0$) 
$$
F(r_1)=0 \hspace{4mm} \Longrightarrow \hspace{3mm} 2m =-\frac{\Lambda}{3} r_1 (r_1^2+1)
$$
so that in particular the function $F$ can be rewritten as
\be
F(r)= \frac{\Lambda}{3} r (r-r_1) (r^2+r r_1 +r_1^2+1) . \label{F1}
\ee
This expression is valid even when $m=0$, in which case we simply set $r_1=0$.
The structure of the metric ---in the first and third terms--- together with the knowledge that $F(r)=0$ has no curvature problem as we have just seen in \eq{invariant}, leads to the idea that $r=0$ and $r=r_1$ ---such that $F(r)=0$--- are horizons, and that the metric is regularly extendible across them. Such an extension is presented in the next section.

Summarizing, the allowed ranges of the used coordinates are:
$$
-\infty < t < \infty , \hspace{3mm} 0\leq \frac{\Lambda}{3} \phi < 2\pi , 
$$
together with
$$
0\leq \theta <\frac{\pi}{2} , \hspace{2mm} \Longleftrightarrow \hspace{2mm}
0\leq x <\infty
$$
and either connected segment of $r$ such that $F(r)\neq 0$.

For completeness and future reference, we also provide the metric \eq{KdS_limit} using the coordinate $x$:
\bea
g=\frac{F(r)\cosh^2 x}{r^2\cosh^2 x +1} \left(dt-\tanh^2 x \, d\phi \right)^2+\frac{\Lambda\tanh^2 x }{3(r^2\cosh^2 x +1)}\left(dt -(r^2+1)d\phi\right)^2 \nonumber\\
+\frac{r^2\cosh^2 x +1}{\cosh^2 x} \left[ \frac{3}{\Lambda}dx^2- \frac{1}{F(r)} dr^2 \right]. \hspace{2cm} \label{gwithx}
\eea

\section{Extension: an alternative derivation using Kerr-type coordinates}
\label{extension}
From the previous discussion we know that the metric \eq{KdS_limit} ---or \eq{gwithx}--- is regularly extendible across $r=0$ and $r=r_1$, the values of $r$ where $F(r)$ vanishes. One can perform such extensions in the usual way, but here we prefer an alternative route which is more effective and allows us to re-derive our spacetime as a limit of the KdS family.

To that end, let us start with the Kerr-de Sitter family given in 
Kerr coordinates (cf.\ e.g.\ \cite{oelz}),
%
\begin{eqnarray}
g_{\mathrm{KdS}}= \frac{-\Delta_r + a^2\sin^2\theta\Delta_{\theta}}{\rho^2}\mathrm{d}v^2 + 2\mathrm{d}r\mathrm{d}v - 2a\sin^2\theta \frac{(a^2+r^2)\Delta_{\theta}-\Delta_r}{\rho^2}\mathrm{d} v\mathrm{d}\varphi
\nonumber
\\
-2a\sin^2\theta \mathrm{d}r\mathrm{d}\varphi + \sin^2\theta \frac{(a^2 + r^2)^2\Delta_{\theta} - a^2\Delta_r\sin^2\theta}{\rho^2}\mathrm{d}\varphi^2
+ \frac{\rho^2}{\Delta_{\theta}}\mathrm{d}\theta^2 ,
\hspace{3mm} 
\label{kerr}
\end{eqnarray}
where
\begin{eqnarray}
\rho^2 &=& r^2 + a^2\cos^2\theta
\,, \label{Deltar}
\\
\Delta_r &=& (r^2 + a^2)\Big(1-\frac{\Lambda}{3}r^2\Big) - 2mr
\,, \nonumber 
\\
\Delta_{\theta} &=& 1 + \frac{\Lambda}{3} a^2\cos^2\theta
\,. \nonumber
\end{eqnarray}
We want to derive the $a\rightarrow \infty$-KdS-limit-spacetime from these coordinates since they are expected to cover a larger domain also when taking the limit and are therefore more suitable to analyze the spacetime. Thus, we introduce new coordinates
\begin{equation*}
v'= av\,, \quad r'= a^{-1} r\,, \quad \varphi'=a^2\varphi
\,,
\end{equation*}
and set
\begin{equation*}
m'=a^{-3}m
\,.
\end{equation*}
Then \eq{kerr} becomes
\begin{eqnarray}
g_{\mathrm{KdS}}= \frac{-\Delta'_r + \sin^2\theta\Delta'_{\theta}}{\rho'^2}\mathrm{d}v'^2 + 2\mathrm{d}r'\mathrm{d}v' - 2\sin^2\theta \frac{(1+r'^2)\Delta'_{\theta}-\Delta'_r}{\rho'^2}\mathrm{d} v'\mathrm{d}\varphi'
\nonumber
\\
-2\sin^2\theta \mathrm{d}r'\mathrm{d}\varphi'  + \sin^2\theta \frac{(1 + r'^2)^2\Delta'_{\theta} -\Delta'_r\sin^2\theta}{\rho'^2}\mathrm{d}\varphi'^2
+ \frac{\rho'^2}{\Delta'_{\theta}}\mathrm{d}\theta^2
\,, 
\label{kerr2}
\end{eqnarray}
where
\begin{eqnarray*}
\rho'^2 &=& r'^2 + \cos^2\theta
\,,
\\
\Delta'_r &=& ( 1+ r'^2)\Big(a^{-2}-\frac{\Lambda}{3} r'^2\Big) - 2m'r'
\,,
\\
\Delta'_{\theta} &=& a^{-2} + \frac{\Lambda}{3}\cos^2\theta
\,.
\end{eqnarray*}
The limit $a\rightarrow \infty$ now leads to a regular metric. Dropping the primes we find
%
\begin{eqnarray}
g&=&\Big(\Delta^{\infty}+ \Sigma^{\infty} \sin^2\theta\Big)  \mathrm{d}v^2 + 2\mathrm{d}v\mathrm{d}r - 2\sin^2\theta
\Big((1+r^2) \Sigma^{\infty}+ \Delta^{\infty}\Big) \mathrm{d} v\mathrm{d}\varphi
\nonumber
\\
&&
-2\sin^2\theta \mathrm{d}r\mathrm{d}\varphi  + \sin^2\theta \Big(\Sigma^{\infty}(1+r^2)^2 + \Delta^{\infty} \sin^2\theta\Big)\mathrm{d}\varphi^2
+\frac{\mathrm{d}\theta^2}{\Sigma^{\infty}}
\,.
\label{KdS_limit2}
\end{eqnarray}
We observe that $\partial_v$ and $\partial_\varphi$ are the two independent and commuting Killing vector fields.
Now the metric is fully regular at $r=0$ and $r=r_1$, where $F(r)$ ---and $\Delta^\infty$--- vanishes, so that the coordinate $r$ takes values in 
$r \in (-\infty,\infty)$. 

On each of the connected intervals for $r$ in which $F(r)\neq 0$ \eq{KdS_limit2} and \eq{KdS_limit} are isometric. 
Thus, \eq{KdS_limit2} is an extension of \eq{KdS_limit}.
The coordinate change that transforms
one into another is 
\begin{equation}
dv  = dt - \frac{r^2+1}{F(r)} d r, \hspace{2cm} 
d \varphi  = d \phi - \frac{dr}{F(r)}. \label{change}
\end{equation}
This implies in particular that $\partial_{\phi} = \partial_{\varphi}$, so
that the axial Killing becomes
$$\eta = \frac{3}{\Lambda} \partial_{\varphi}.$$
Hence, the ranges of the coordinates are
$$
-\infty < r < \infty , \hspace{3mm} -\infty < v < \infty , \hspace{3mm} 0\leq \frac{\Lambda}{3}\varphi < 2\pi , \hspace{3mm}
0\leq \theta <\frac{\pi}{2}  \hspace{2mm} (\Longleftrightarrow \hspace{2mm}
0\leq x <\infty).
$$

For completeness, again we provide the form of this extended metric using the coordinate $x$: 
\begin{align}
g=&\frac{1}{r^2\cosh^2 x +1} \left(F(r)\cosh^2 x +\frac{\Lambda}{3} \tanh^2 x\right)dv^2+\left( \frac{3}{\Lambda}\right) \frac{r^2\cosh^2 x +1}{\cosh^2 x} dx^2 \nonumber\\
&-\frac{2}{r^2\cosh^2 x +1} \left(F(r)\sinh^2 x +\frac{\Lambda}{3} (r^2+1) \tanh^2 x\right) dv d\varphi -2\tanh^2 x\,  dr d\varphi \nonumber\\
&+2 dv dr +\frac{\tanh^2 x}{r^2\cosh^2 x +1} \left(F(r)\sinh^2 x +\frac{\Lambda}{3} (r^2+1)^2 \right) d\varphi^2.\hspace{1cm} \label{gwithx2}
\end{align}
\subsection{Generalized Kerr-Schild form}\label{Kerr-Schild}
It may be observed that the extended metric \eq{KdS_limit2} is explicitly {\em linear} 
(or, strictly speaking, {\em affine}) 
on the parameter $m$. Collecting the terms proportional to $m$ we can rewrite \eq{KdS_limit2} as
\bean
g&=&\frac{\Lambda}{3} \left(r^2+ \sin^2\theta\right)  \mathrm{d}v^2 + 2\mathrm{d}v\mathrm{d}r -2\sin^2\theta \mathrm{d}r\mathrm{d}\varphi 
\\
&&
-\frac{2\Lambda }{3}(r^2+1) \sin^2\theta \mathrm{d} v\mathrm{d}\varphi + \frac{\Lambda }{3}(r^2+1) \sin^2\theta \mathrm{d}\varphi^2
+\frac{\mathrm{d}\theta^2}{\Sigma^{\infty}}
\\
&&+\frac{2mr}{r^2+\cos^2\theta} \left(dv^2 -2\sin^2\theta dv d\varphi +\sin^4\theta d\varphi^2 \right)
\,.
\eean
We know that our metric has positive constant curvature when $m=0$, and in that case 
$g|_{m=0}:=\tilde g$ with $\tilde g$ given by the first two lines of the previous expression 
\bea
\tilde g  &=&\frac{\Lambda}{3} \left(r^2+ \sin^2\theta\right)  \mathrm{d}v^2 + 2\mathrm{d}v\mathrm{d}r -2\sin^2\theta \mathrm{d}r\mathrm{d}\varphi 
\nonumber
\\
&&
-\frac{2\Lambda }{3}(r^2+1) \sin^2\theta \mathrm{d} v\mathrm{d}\varphi + \frac{\Lambda }{3}(r^2+1) \sin^2\theta \mathrm{d}\varphi^2
+\frac{\mathrm{d}\theta^2}{\Sigma^{\infty}} .
\label{dS1}
\eea
Therefore, $\tilde g$ is the 
 metric in (possibly a portion of) de Sitter spacetime. One can find the explicit coordinate transformation to standard forms of de Sitter in standard coordinates, and this is given in the Apendix, but this change is complicated and not helpful 
in clarifying the global structure of our spacetime because the horizon structure changes dramatically when $m$ vanishes, as we are going to discuss.

Our metric takes the explicit generalized Kerr-Schild form based on the de Sitter seed metric (\ref{dS1})
$$
g=\tilde g +2 Q \ell \otimes \ell
$$
with 
$$Q=\frac{mr}{r^2+\cos^2\theta}$$
and where $\ell$ is the null one-form (both in $g$ and $\tilde g$) given by 
\be
\ell := dv -\sin^2\theta d\varphi = dv -\tanh^2 x \, d\varphi \label{ell}.
\ee
This is actually one of the two principal null directions of the Petrov type-D metric. 

Using the coordinate $x$ we can write for the de Sitter seed metric
\bea
\tilde g=\frac{\Lambda}{3(r^2\cosh^2 x +1)} \left(r^2(r^2+1) \cosh^2 x + \tanh^2 x\right)dv^2+\left( \frac{3}{\Lambda}\right) \frac{r^2\cosh^2 x +1}{\cosh^2 x} dx^2 \nonumber\\
-\frac{2\Lambda}{3} (r^2+1)  \tanh^2 x \, dv d\varphi -2\tanh^2 x\,  dr d\varphi 
+2 dv dr +\frac{\Lambda}{3} (r^2+1)  \tanh^2 x \, d\varphi^2 \hspace{3mm} \label{dS2}
\eea
and also
$$
Q=\frac{mr\cosh^2 x}{r^2\cosh^2 x+1} \, .
$$

\section{Killings' properties and trapped surfaces}\label{TrappedSurfaces}

The one-form $dr$ is such that ($g^{\sharp}$ denotes the inverse of the metric $g$ )
$$
g^\sharp (dr,dr) =g^{rr}= -\Delta^\infty =-\frac{F(r)}{r^2+\cos^2\theta} =-\frac{F(r)\cosh^2 x}{r^2\cosh^2 x +1}
$$
hence the hypersurfaces $r=$const.\ are timelike between the two values $r=0$ and $r=r_1$ (where $F(r)<0$), spacelike outside that interval, and null at $r=0$ and $r=r_1$. The extension of the original spacetime has been achieved precisely across these two null hypersurfaces.

In order to check whether or not the null hypersurfaces $r=0$ and $r=r_1$ are Killing horizons, we compute the norm of a general Killing vector
$$
\beta \partial_v +\alpha \partial_\varphi
$$
with $\alpha,\beta\in \mathbb{R}$ and we obtain that (i) it vanishes on $r=0$ if $\beta =\alpha$, so that $r=0$ is a Killing horizon for the Killing vector
\be
\xi = \alpha ( \partial_v +\partial_{\varphi}  ) , \quad \quad \alpha \neq 0 \label{killing1} ;
\ee
and (ii) it vanishes on $r=r_1$ for $\beta =\alpha (1+r_1^2)$, hence $r=r_1$ is a Killing horizon for the Killing vector
\be
\zeta = \alpha ( (1+ r_1^2) \partial_v + \partial_{\varphi} ) \quad \quad \alpha \neq 0\label{killing2} .
\ee
We keep the constant $\alpha$ arbitrary to adjust the physical dimensions 
when computing the surface gravity of the horizons later on. We nevertheless
fix, for concreteness, $\alpha >0$ and
determine the causal orientation of the Killing vectors. From the structure of the metric the vector field $\partial_r$ is obviously 
everywhere null ($g_{rr}=0$) and nowhere zero. 
Now, $\xi$ is also null on the horizon $r=0$, but their scalar product is positive
$$
g(\partial_r,\xi)=\alpha \cosh^{-2} x>0
$$ 
implying that $\xi $ has opposite causal orientation than $\partial_r$
 on this horizon. Similarly, we have
$$
g(\partial_r,\zeta)=\frac{\alpha ( r_1^2\cosh^2 x+1)}{\cosh^{2} x} >0
$$
so that $\zeta$ also has opposite causal orientation
than  $\partial_r$ on the Killing horizon $\{ r=r_1 \}$.

Actually, one can compute the causal character of the orbits of the $G_2$ group generated by the two Killing vectors $\{\partial_v,\partial_\varphi\}$, which is ruled by the determinant of their scalar products, and the result is
$$
\det\left( \begin{array}{cc}
g_{vv} & g_{v\varphi} \\
g_{\varphi v} & g_{\varphi\varphi} 
\end{array}
\right) =\frac{\Lambda}{3}\sin^2\theta\cos^2\theta\,  F(r)=
\frac{\Lambda}{3}\frac{\sinh^2 x}{\cosh^4 x} F(r)
$$
so that these orbits are timelike between $r=0$ and $r=r_1$, spacelike outside that interval, and null at the Killing horizons. 

With regard to the causal character of the Killing vectors themselves, the norm of $\partial_\varphi$ 
reads
$$
g(\partial_\varphi,\partial_\varphi)= 
\frac{\tanh^2 x}{r^2\cosh^2 x +1} \left(F(r)\sinh^2 x +\frac{\Lambda}{3} (r^2+1)^2 \right)
$$
and thus, outside the axis, this Killing vector is spacelike wherever 
$$
F(r)\sinh^2 x +\frac{\Lambda}{3} (r^2+1)^2  >0 .
$$
This holds in the regions with $F(r)>0$. However, in the region between $r=0$ and $r=r_1$, where $F(r)<0$, one can always find values of $x$ such that $\partial_\varphi$ is timelike. In those zones, therefore, the integral curves of this Killing vector are closed timelike curves. 

The hypersurfaces $r=r_0=$const.\ are foliated by 2-dimensional surfaces with $v=v_0=$const.\  Their first fundamental forms read
$$
\gamma |_{v_0,r_0} =\left( \frac{3}{\Lambda}\right) \frac{r_0^2\cosh^2 x +1}{\cosh^2 x} dx^2
+\frac{\tanh^2 x}{r_0^2\cosh^2 x +1} \left(F(r_0)\sinh^2 x +\frac{\Lambda}{3} (r_0^2+1)^2 \right) d\varphi^2
$$
so that they are all spacelike surfaces outside the region with closed timelike curves, in particular whenever $F(r)\geq 0$. Their mean curvature one-form can be computed by e.g. using the formulas in \cite{S2002} and the result is
\be
H = \frac{1}{2} \frac{F'(r_0)\sinh^2 x +\frac{4\Lambda}{3} r_0(r_0^2+1)}{F(r_0)\sinh^2 x +\frac{\Lambda}{3} (r_0^2+1)^2}\, dr \label{mean}
\ee
whose norm is
$$
g^\sharp (H,H)=-\frac{1}{4} \left(\frac{F'(r_0)\sinh^2 x +\frac{4\Lambda}{3} r_0(r_0^2+1)}{F(r_0)\sinh^2 x +\frac{\Lambda}{3} (r_0^2+1)^2}\right)^2 \frac{F(r_0)\cosh^2 x}{r_0^2\cosh^2 x +1}.
$$
Therefore, in the region with $F(r)<0$ ---but outside the zone with closed timelike curves--- these spacelike surfaces are untrapped. On the other hand, in the regions with $F(r)>0$ the mean curvature vector is timelike, and thus these surfaces are trapped. In these regions the causal orientation of $H$ is ruled by the sign of the numerator in (\ref{mean}). Fixing $m>0$ for definiteness, we know that  
$F'(r)<0$ for $r<r_1<0$, and thus the mentioned numerator is negative for $r_0< r_1$. Similarly, $F'(r)>0$ for $r>0$, and thus the sign of the numerator is positive for $r_0>0$. Therefore, using that $dr(\partial_r)=1$, the causal orientation of $H$ is the same as that of the null vector field $\partial_r$ for $r_0<r_1$, and the opposite for $r_0>0$. Hence, if $\partial_r$ is taken to be future pointing, then these surfaces are future-trapped for $r_0<r_1$ and past-trapped for $r_0>0$. The same reasoning, using $F(r_1)=F(0)=0$ and $F'(r_1)<0$ and $F'(0)>0$, proves that these surfaces are marginally trapped on the Killing horizons. The case with $m<0$ can be treated analogously.

It must be remarked that these trapped surfaces are {\em non-compact} with infinite area. This easily follows from their area calculation  
$$
\mbox{Area}(r_0,v_0)=2\pi \left ( \frac{3}{\Lambda} \right )^{\frac{3}{2}} \int^\infty_0 \frac{\sinh x}{\cosh^2 x}\left(F(r_0)\sinh^2 x +\frac{\Lambda}{3} (r_0^2+1)^2 \right)^{1/2} dx 
$$
which behaves as 
$$
\mbox{Area}(r_0,v_0)\sim  2\pi \left ( \frac{3}{\Lambda}
\right )^{\frac{3}{2}}  F(r_0) x \hspace{1cm} \mbox{when}\, \, \,  x\rightarrow \infty .
$$
Still, the area does not diverge for these surfaces on the two Killing horizons, where $F=0$, and this will be discussed in the next section.

Concerning the other Killing vector field $\partial_v$, its norm reads
$$
g(\partial_v,\partial_v)=\frac{1}{r^2\cosh^2 x +1} \left(F(r)\cosh^2 x +\frac{\Lambda}{3} \tanh^2 x\right)
$$
hence $\partial_v$ is spacelike everywhere in the regions with $F(r)>0$, as well as on the horizons $r=0$ and $r=r_1$ outside the axis of symmetry. At the intersection of the horizons with the axis $\partial_v$ becomes null. In the region between the two horizons this Killing is timelike near the axis and for large $x$, and it is spacelike in the region with
$$
0< -F(r) < \frac{\Lambda}{3}\frac{\sinh^2 x}{\cosh^4 x} \leq \frac{\Lambda}{12}\, .
$$

By noting that  $dv$ is orthogonal to 
the null field $\partial_r$ 
and that $\ell = g(\partial_r, \cdot)$ as given in (\ref{ell}) 
is linearly independent to $dv$ outside the axis (on the axis they
are linearly dependent -- in fact the same --  
since the one-form
$\tanh^2 x d \varphi$ can be seen to extend smoothly to 
zero on the axis of symmetry)
we deduce that $dv$ is spacelike away from the axis
and thus the hypersurfaces $v=$const.\ are
timelike everywhere outside the axis and null on the axis of symmetry.


\section{The Killing horizons; angular momentum}\label{KillingHorizons}
By definition, Killing horizons are foliated by 
marginally outer trapped surfaces (MOTSs): any spacelike cut on a Killing horizon is a MOTS. In our case, as proven in the previous section, the spacelike cuts of these horizons are actually marginally trapped surfaces (MTS), as the causal orientation of their null mean curvature vector is the same at every point of the horizon. To understand the intrinsic geometry of these horizons we compute their first fundamental forms $\gamma$. For the case of the 
$\{r=0\}$ horizon we obtain
\be
\gamma |_{r=0}= \frac{3}{\Lambda} d\theta^2 +\frac{\Lambda}{3} \sin^2\theta (dv-d\varphi)^2
\label{1FF}
\ee
which is obviously degenerate. The above metric looks like the standard one for round spheres. However, we have to remember that the spacetime has a curvature singularity at the values \eq{singularity}, and therefore the above is the metric for topological open disks with $\theta\in [0,\pi/2)$ and constant Gaussian curvature. Still, their area is finite (and equal to $6\pi /\Lambda$). It follows that this Killing horizon is foliated by {\em non-compact} MTSs with finite area.

We now compute the surface gravity $\kappa_0$ associated to this Killing
horizon. For definiteness, assume that $\partial_r$
is future pointing (the other case is similar), so that, from the discussion of the previous section, the MTSs on the horizon are past trapped if $m>0$ 
(its transverse future null expansion is positive). Then, the future pointing Killing generator of the
Killing horizon $\{r=0\}$
is $-\xi$ with $\xi$ given in (\ref{killing1}) 
 and the corresponding surface gravity is
$$
(-\xi^\rho) \nabla_\rho (-\xi^\mu) |_{r=0}= \kappa_0 (-\xi^\mu )|_{r=0} 
\quad  \Longleftrightarrow \quad \nabla_\mu g(\xi ,\xi) |_{r=0} = 2\kappa_0 \xi_\mu |_{r=0}.
$$
Taking into account that
$$
g(\xi,\xi) = \alpha^2 \left ( g_{vv} +2 g_{v\varphi} + g_{\varphi\varphi} \right ) =
\frac{\alpha^2}{r^2\cosh^2 x +1} \left( \frac{F(r)}{\cosh^2 x} + \frac{\Lambda}{3} r^4 \tanh^2 x \right)
$$
a straightforward calculation gives 
\be
\kappa_0 = \alpha F'(0)/2= \alpha m. \label{kappa0}
\ee
We now chose $\alpha$ to adjust correctly the physical dimensions. For 
Killing vectors with the usual physical dimensions of $L^{-1}$ (as e.g.
the standard time translation in Schwarzschild), the surface gravity has dimensions of $L^{-1}$. From the discussion on dimensions
in Section \ref{Basics} and (\ref{change}) a natural choice is
 $\alpha = \sqrt{\frac{3}{\Lambda}}$ since then
the Killing
\begin{align*}
\xi = \sqrt{\frac{3}{\Lambda}} \left ( \partial_v + \partial_{\varphi} 
\right )
\end{align*}
has the appropriate  dimension $[\xi] = L^{-1}$. The surface gravity of the
Killing horizon $\{ r= 0\}$ for this choice of Killing generator 
is
 hence  $\kappa_0 = \sqrt{\frac{3}{\Lambda}} m$.
The Killing horizon is degenerate when $m=0$ as expected, since
in this limit the two Killing horizons at $\{ r=0 \} $ and $\{ r=r_1\}$
merge into one.
If the Killing horizon lies in a domain where $\partial_r$ is past pointing, the
future generator is now $\xi$ and the corresponding surface gravity is
$\kappa_0 = - \sqrt{\frac{3}{\Lambda}} m$.

Concerning the other Killing horizon at $r=r_1$, the degenerate induced metric reads
\bean
\gamma |_{r=r_1}&=& \frac{3}{\Lambda} (r_1^2 +\cos^2\theta)\frac{d\theta^2}{\cos^2\theta}+\frac{\Lambda}{3} \frac{\cos^2\theta\sin^2\theta}{r_1^2+\cos^2\theta}\left(dv-(1+r_1^2)d\varphi\right)^2 \\
&=& \frac{3}{\Lambda}(r_1^2\cosh^2 x+1)\frac{dx^2}{\cosh^2 x} +
\frac{\Lambda}{3} \frac{\tanh^2 x}{r_1^2\cosh^2 x+1}\left(dv -(1+r_1^2) d\varphi \right)^2 \, .
\eean
Surfaces with this metric do not have constant curvature. Their Gaussian curvature reads
$$
K=\frac{\Lambda}{3}\, \frac{1+3r_1^2\cosh^2 x +3r_1^2(1+2r_1^2)\cosh^4 x -r_1^4\cosh^6 x}{(r_1^2\cosh^2 x+1)^3}
$$
which is positive at the axis $x=0$ and then decreases very rapidly approaching the negative value $-\Lambda/(3r_1^2)$ as $x\rightarrow \infty$. Observe that the distances from the center at the axis along the``radial'' curves with tangent $\partial_x$ have no finite bound, and they tend to $+\infty$ as $x\rightarrow \infty$. Therefore, these surfaces are non-compact, despite the fact  that they again have a finite area, given by 
$6 \pi (1+ r_1^2)/\Lambda$.
This can happen because the length along the circles with constant $x=x_0$ reaches a maximum and then decreases, as $x_0$ increases, indefinitely tending to zero as $x_0\rightarrow \infty$.
This Killing horizon is thus foliated by non-compact MTSs with finite area too.

Assuming that $\partial_r$ is future pointing ---and for $m>0$ say---, the MTSs are future trapped and
the surface gravity $\kappa_1$  of 
the horizon $\{ r=r_1 \} $ 
is now defined by 
$$
(-\zeta^\rho ) \nabla_\rho (-\zeta^\mu ) |_{r=r_1}= \kappa_1 (-\zeta^\mu )|_{r=r_1} 
\quad \Longleftrightarrow \quad  \nabla_\mu g(\zeta ,\zeta) |_{r=r_1} = 2\kappa_1 \zeta_\mu |_{r=r_1}
$$
with $\zeta$ given in 
(\ref{killing2}) with the same choice
$\alpha = \sqrt{\frac{3}{\Lambda}}$ so that 
$[ \zeta ] = L^{-1}$. Then, on using 
$$
g(\zeta,\zeta)  =
\frac{3}{\Lambda (r^2\cosh^2 x +1)} \left( \frac{F(r)}{\cosh^2 x}(r_1^2\cosh^2 x +1)^2 + \frac{\Lambda}{3} \tanh^2 x (r^2-r_1^2)^2 \right),
$$
the following value 
\be
\kappa_1 = \sqrt{\frac{3}{\Lambda}}
F'(r_1)/2 = 
\sqrt{\frac{3}{\Lambda}} \left ( 
-m+\frac{\Lambda}{3} r_1^3 \right ) = 
\frac{1}{2} \sqrt{
\frac{\Lambda}{3}} r_1 (3 r_1^2+1) \label{kappa1}
\ee 
is obtained. If 
$\partial_r$ is past-pointing, the surface gravity with respect to
the future generator $\zeta$ is the opposite of (\ref{kappa1}).

Note that the sign of $\kappa_1$ is always opposite to the sign of $\kappa_0$.
For instance, and assuming $\partial_r$ to be future,
if $m>0$, then $\kappa_0 >0$ while $\kappa_1< 0$, and vice versa for $m<0$. 

\subsection{Mass and angular momentum}
The geometry of the horizons can be used to interpret the parameters
in the metric. The cosmological constant $\Lambda$ fixes the overall scale, so 
the metric has only one essential parameter $m$.  In order to interpret
$m$ physically it makes sense to compute the angular momentum
of each of the Killing horizons. Recall that the angular momentum of an axially
symmetric Killing horizon ${\cal H}$ with axial Killing vector $\eta$
is defined as follows: take a cross section $\Sigma$
of the horizon  and assume that $\eta$ is tangent to
$\Sigma$ everywhere. Take a basis $\{n,k\}$ 
of null normal vector fields to $\Sigma$ where $n$ is the null generator of ${\cal H}$ and fix partially their boost freedom 
by  $g(n, k) =-2$. The total
angular
momentum of the section $\Sigma$ is defined by (see e.g. \cite{Jaramillo})
\begin{align}
J(\Sigma) = -\frac{1}{8 \pi} \int_{\Sigma} s(\eta) dS_{\Sigma}\label{Jdef}
\end{align}
where $dS_{\Sigma}$ is the metric measure of $\Sigma$, and given the null basis the connection one-form $s$ is defined, on tangent vector fields $X$ to $\Sigma$,
by the expression
\begin{align*}
s (X)= - \frac{1}{2} g(k, \nabla_X n) \, .
\end{align*}
When $\H$ has compact cross sections, it is 
well-known that $J(\Sigma)$ is independent of the cross section (see e.g.
\cite{MarsTotally}). This result extends to horizons with non-compact cross
sections as long as the topology of the horizon is 
$\mathbb{R} \times
\Sigma$. This follows easily from the transformation law of 
$s$ under change of cross section, first obtained in \cite{Ashtekhar}. 
Using the notation in 
\cite{MarsTotally} a cross section $\Sigma'$ defined
by a graph function $f: \Sigma \rightarrow \mathbb{R}$ has one-form connection
$s[\Sigma']$ given by
\begin{align*}
s [\Sigma'] = s[\Sigma_v] + \kappa \, d f
\end{align*}
where $\kappa$ is the surface gravity of the horizon null generator $n$
and $\Sigma_{v} := \{v\} \times \Sigma$.
Assume that each $\Sigma_v$ is axially symmetric. Then,
the section $\Sigma'$ will be axially symmetric (i.e. with 
$\eta$ tangent to $\Sigma'$) iff $\eta(f)=0$, and the independence
of $J(\Sigma)$ with respect to the cross section is immediate
from its definition. 

Thus, we may consider any cross section $\{ v=\mbox{const}, r\in\{0,r_1\}\}$
to compute the angular momentum of the horizons. We
concentrate on the horizon $\{ r=r_1 \}$ and denote its
angular momentum by $J_1$.
Noting that the null generator is then $n =\zeta$ as given in (\ref{killing2}) with $\alpha =\sqrt{3/\Lambda}$, the second null normal vector field $k$ is
\begin{align*}
k = -\sqrt{\frac{3}{\Lambda}} \frac{\cosh^2 x}{(1+ r_1^2 \cosh^2 x)} \left ( 
\frac{\sinh^2 x}{1+r_1^2} \partial_v 
+ \frac{2 \Lambda}{3}  \partial_r + 
\frac{1 + (1+2 r_1^2) \cosh^2 x}{(1+r_1^2)^2} 
\partial_{\varphi}\right ).
\end{align*}
A straightforward calculation provides 
$$
s(\eta) = \frac{1}{2} \frac{\sinh^2 x}{\left ( 1+r_1^2\cosh^2x \right )^2} r_1 \left[(3r_1^2 +1) r_1^2 \cosh^2 x +r_1^2 -1 \right].
$$
Applying then directly the definition (\ref{Jdef}) and performing the integral, the result
is simply
\begin{align}
J_1 = \frac{9 m}{2\Lambda^2}.
\label{angular}
\end{align}
The angular-momentum of the
Killing horizons in the Kerr-de Sitter spacetime (of mass parameter
$\tilde{m}$ and specific angular momentum $a$) reads (see e.g \cite{Cho})
\begin{align*}
J = \frac{a \tilde{m}}{\left ( 1+ \frac{\Lambda}{3}a^2 \right )^2}
\end{align*}
Taking the limit $a \rightarrow \infty$  as discussed in Sect. \ref{extension}
one finds $9m /\Lambda^2$, which is {\it twice} the result for the
limit spacetime. Thus, the angular momentum 
is a discontinuous function of $a$  as $a \rightarrow \infty$. Such discontinuous
behaviour  also occurs for other quantities associated to the horizon, for
instance the area of its cross sections. 
The area of a Killing horizon $r=\hat r$ in Kerr-de Sitter is
\begin{equation*}
\mbox{Area}(\hat r,v_0) = \frac{4 \pi (\hat r^2 + a^2 )}{1+ \frac{\Lambda}{3}a^2}.
\end{equation*}
where $\hat r$ is a root of $\Delta_r=0$. For $a$ sufficiently large 
and $\tilde{m}= a^3 m$ with fixed $m$, it is easy
to prove that this polynomial has exactly two roots $r_\pm$ with behaviour
$r_{+} = O(1/a)$ and $r_{-} = a r_1 + O(1/a)$. 
Hence, the area
of the horizon at $r_{+}$  is 
$$
\mbox{Area}_{KdS}(r_+,v_0)= \frac{12 \pi}{\Lambda} + o(a^{-2})
$$ 
whose limit is not $6 \pi /\Lambda$. Similarly, the area of the horizon at 
$r_{-}$ is 
$$
\mbox{Area}_{KdS}(r_-,v_0)=\frac{12 \pi (1+ r_1^2)}{\Lambda} + o(a^{-2})
$$ 
and its limit is 
again different from $6 \pi (1+r_1^2)/\Lambda$. In both cases 
the limit of the areas is exactly twice the area
at the limit. This reflects the fact that, while $\theta$ takes values
in $(0,\pi)$ in Kerr-de Sitter, its range has dropped to $(0,\pi/2)$
in the $a \rightarrow \infty$ limit spacetime. A similar factor $2$ 
was 
found in the behaviour of the total angular momentum at the horizon
$\{r = r_1\}$,
so one might think that this is a general fact. However, this is
not the case. While the total angular momentum of any horizon in the Kerr-de Sitter
spacetime has a limit $9 m /\Lambda^2$ when $a \rightarrow \infty$, the 
total angular momentum
of the horizon $\{r=0\}$ in the
$a\rightarrow \infty$-KdS-limit-spacetime
is infinity. We emphasize, however, that this value is
of little physical significance
because the horizon $\{r=0\}$ touches the 
spacetime singularity at $\{r =0,\theta=\pi/2\}$.

In any case, we can conclude from the previous discussion
that $m$ measures directly the total rotation of the spacetime. 
Moreover, we know that the limit $m=0$ is the de-Sitter spacetime, so $m$
must also be a measure of the total mass/energy of the spacetime. 
However, in the case of spacetimes with positive cosmological constant, 
it is not yet clear what this total mass should be.
There have been
recent attempts to define the Bondi energy associated to 
closed surfaces at $\scri$ in spacetimes with positive
cosmological constant \cite{Szabados}, \cite{Chrusciel}. 
However, the definitions are either difficult to compute
explicitly \cite{Szabados}, or provide expressions for which few general
properties are known \cite{Chrusciel} (e.g. whether they vanish identically for
arbitrary closed surfaces at $\scri$ in the de Sitter space, or whether
restrictions on the ``cuts'' must be imposed to make
the quantity physically meaningful). This prevents us from
reaching any conclusion concerning the relation between $m$ and
the total mass. The only thing we can say at this point is that, 
given that energy has dimensions of length  any sensible definition
of Bondi energy must yield
\begin{align*}
M = c \frac{m^{\mu}}{\Lambda^{\frac{1}{2} + \mu}}
\end{align*}
where $c$ is a constant that may depend on the cut and $\mu$ is a fixed
positive constant (so that $m \rightarrow 0$ yields zero mass).

\section{Further extensions: the global structure}\label{extensions2}
Equation \eq{killing1}, as well as \eq{1FF}, both suggest to introduce a new angular coordinate by
\begin{equation}
\Phi = \varphi -  v
\end{equation}
so that the axial Killing vector still remains invariant $\partial_\varphi 
=\partial_\Phi$.
Then \eq{KdS_limit2}  becomes
\begin{eqnarray}
g
&=&\Big(\cos^4\theta \Delta^{\infty}+ r^4\sin^2\theta \Sigma^{\infty}\Big)  \mathrm{d}v^2 
+ 2\cos^2\theta\mathrm{d}  v \mathrm{d}r
\nonumber
\\
&&
-\frac{4mr}{r^2+\cos^2\theta} \sin^2\theta \cos^2\theta \mathrm{d} v\mathrm{d}\Phi
-2\sin^2\theta \mathrm{d}r\mathrm{d}\Phi
\nonumber
\\
&&
+ \sin^2\theta \Big((1+r^2)^2 \Sigma^{\infty}+ \sin^2\theta \Delta^{\infty}\Big)\mathrm{d}\Phi^2
+\frac{\mathrm{d}\theta^2}{\Sigma^{\infty}} \label{KdS_limit3}
\end{eqnarray}
or alternatively 
\begin{align}
g= & \frac{1}{\cosh^2 x (r^2\cosh^2 x +1)}\left(F(r)+\frac{\Lambda}{3} r^4 \sinh^2 x  \right) dv^2
+\frac{2}{\cosh^2 x} dv dr \nonumber 
\\
&+\left( \frac{3}{\Lambda}\right) \frac{r^2\cosh^2 x +1}{\cosh^2 x} dx^2 -2\tanh^2 x \, dr d\Phi -\frac{4mr }{r^2\cosh^2 x+1}\tanh^2 x\,  dv d\Phi \nonumber
\\
& +\frac{\tanh^2 x}{r^2\cosh^2 x +1} \left(\frac{\Lambda}{3} (r^2+1)^2+F(r) \sinh^2 x \right)d\Phi^2
\end{align}
while the de Sitter forms (\ref{dS1}) and (\ref{dS2}) read now
\begin{eqnarray}
\tilde g  &= & \frac{\Lambda}{3} r^2\cos^2\theta \mathrm{d}v^2 
+ 2\cos^2\theta\mathrm{d}  v \mathrm{d}r
-2\sin^2\theta \mathrm{d}r\mathrm{d}\Phi
\nonumber
\\
&&
+ \frac{\Lambda}{3}(1+r^2) \sin^2\theta\mathrm{d}\Phi^2
+\frac{3} {\Lambda}(r^2 +\cos^2\theta)\frac{\mathrm{d}\theta^2}{\cos^2\theta}
\label{deSitter_EF} 
\\
&=&\frac{\Lambda}{3} \frac{r^2}{\cosh^2 x} \mathrm{d}v^2 +\frac{2}{\cosh^2 x} dv dr 
-2\tanh^2 x \, dr d\Phi\nonumber
\\
&&+\frac{\Lambda}{3} (r^2+1) \tanh^2 x \, d\Phi^2+\left( \frac{3}{\Lambda}\right) \frac{r^2\cosh^2 x +1}{\cosh^2 x} dx^2 \label{dS3}
\end{eqnarray}
and the one-form \eq{ell} becomes
\begin{equation}
\ell = \cos^2\theta \mathrm{d}v - \sin^2\theta\mathrm{d}\Phi = \frac{1}{\cosh^2 x}\left(dv - \sinh^2 x d\Phi \right). \label{ell2} 
\end{equation}

Using the new angular coordinate, the Killing vector fields \eq{killing1} and \eq{killing2} that become null at the horizons read now
\be
\xi = \sqrt{\frac{3}{\Lambda}} 
\partial_{v}, \hspace{1cm} \zeta = \sqrt{\frac{3}{\Lambda}} 
\left ( (1+r_1^2) \partial_v -r_1^2 
\partial_\Phi \right ) .\label{killings}
\ee
We know that $\xi$ is, on the Killing horizon $\{r=0\}$, a null geodesic vector field tangent to the generators of the horizon with $\xi^\rho\nabla_\rho \xi^\mu = - \sqrt{\frac{3}{\Lambda}} m \xi^\mu$ (due to \eq{kappa0}). Therefore, the affinely parametrized geodesic vector field tangent to the generators of the horizon is $\hat{\xi}:=e^{mv}\xi$. But this immediately implies that an affine parameter along these null geodesics is proportional to $e^{-mv}$, and thus they are incomplete for $v\rightarrow \infty$ if $m>0$, and for $v\rightarrow -\infty$ if $m<0$. Observe that this means, due to the orientation of $\xi =\partial_v$, 
that if $\partial_r$ is future pointing they are incomplete to the {\em past} in the former case, and to the future in the latter, and vice versa
if $\partial_r$ is past pointing.

An analogous argument using $\kappa_1$ proves that the null geodesics tangent to
$\zeta$ at the other horizon $r=r_1$ are incomplete as $v\rightarrow -\infty$ if $m>0$ (that is, future incomplete), or as $v\rightarrow \infty$ if $m<0$ (past incomplete) whenever $\partial_r$ is future, and vice versa when
$\partial_r$ is past. This strongly indicates that all the horizons will in 
fact be bifurcate (as long as $m \neq 0$) in a further extended spacetime. 

By thinking of a conformal diagram in the $(v,r)$ plane, it seems that the horizons satisfy (set $m>0$ and $\partial_r$ future pointing,
the other cases are similar)
\begin{itemize}
\item $\{ r=0\} $ is a Killing horizon complete to the future, and thus touching $\scri^+$, but incomplete to the past;
\item $\{ r=r_1 \}$ is a Killing horizon complete to the past, so that approaches $\scri^-$, but incomplete to the future.
\end{itemize}
From this we also guess that all null geodesics are going to be complete to the past if they have $r\leq r_1$, but incomplete to the past if they have $r>r_1$. And all null geodesics will be complete to the future if they have $r\geq 0$, but incomplete if they live in $r<0$. To check that these statements are correct, we analyze the null geodesics at the axis. As is known \cite{MS}, the regular axis of an axial Killing vector is a totally geodesic timelike 2-dimensional submanifold, hence the intrinsic geodesics of the axis as a submanifold are actually geodesics of the full spacetime. 
The first fundamental form on the axis $x=0$ reads simply
\be
\gamma|_{x=0} = 2dv dr +\frac{F(r)}{r^2+1} dv^2 \label{axis}
\ee
and the null field $\partial_r$ is affinely parametrized (this follows
immediately from the fact that its metrically related  
one-form $dv$ is null and exact). Thus, all the null geodesics
at constant $v$ are complete. For the remaining affinely
parametrized geodesics, 
and letting aside the two null geodesics at $r=0$ and $r=r_1$ 
already discussed,  they are the integral curves of the null
vector field
$$
N:= \frac{2(r^2+1)}{F(r)} \partial_v -\partial_r \,. \label{k}
$$
This null vector field is affinely parametrized  because it is null,
integrable and its scalar product with the Killing field
$\partial_v$ is constant and non-zero.
Moreover 
$g(N,\partial_r) = 2 (r^2 +1) /F(r)$, so 
this vector field has opposite causal orientation to
$\partial_r$  in the regions with $F(r)>0$, and the same orientation
in the region with $F(r)<0$. Therefore, keeping as before $m>0$ and
$\partial_r$ future directed for concreteness, the following possibilities arise (here $r_0$ is the initial value of $r$ at the null geodesics):
\begin{itemize}
\item if $r_0<r_1$ then these null geodesics are 
\begin{itemize}
\item complete to the past reaching $\scri^-$
\item incomplete to the future reaching $r=r_1$ as $v\rightarrow -\infty$
\end{itemize}
\item if $r_1<r_0<0$ then they are
\begin{itemize}
\item incomplete to the past reaching $r=0$ as $v\rightarrow \infty$
\item incomplete to the future reaching $r=r_1$ as $v\rightarrow -\infty$
\end{itemize}
\item if $r_0>0$ then they are
\begin{itemize}
\item incomplete to the past reaching $r=0$ when $v\rightarrow \infty$
\item complete to the future reaching $\scri^+$
\end{itemize}
\end{itemize}
This agrees with our guess, and a similar analysis can be performed for more complicated geodesics.

\begin{figure}[!ht]
\begin{center}
\includegraphics[height=6cm]{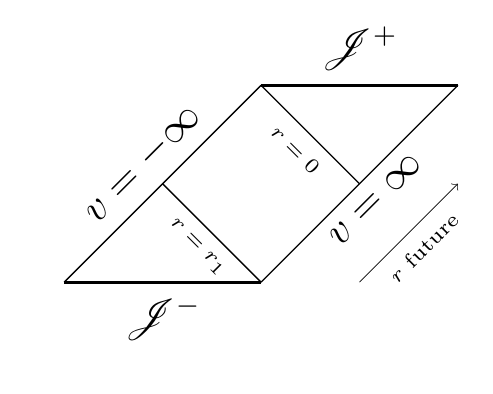}
\end{center}
\caption{\footnotesize{Schematic conformal diagram of the axis $x=0$. As usual, null lines are drawn at 45$^o$ and the future direction is upwards. In this diagram we assume that the parameter $m$ is positive, and that $\partial_r$ is a null vector field pointing to the future too. The two Killing horizons intersect the axis at the null geodesics labeled as $r=r_1$ and $r=0$. Notice that $r_1<0$. Past null infinity, where $r\rightarrow -\infty$, is denoted by $\scri^-$, while future null infinity ($r\rightarrow \infty$) is $\scri^+$. We know from the considerations on \cite{mpss2} that both $\scri$'s have the topology of $\mathbb{S}^3\setminus \{p\} =\mathbb{R}^3$. Null geodesics traveling along the vector field $\partial_r$ extend all the way from $\scri^-$ to $\scri^+$ and are complete. The second family of null geodesics ---parallel, and including, the lines marking the horizons--- are incomplete to the future if they have $r<0$, while they are incomplete to the past if they have $r>r_1$. Therefore, this metric is extendible across the two borders marked by $v=\pm \infty$.}}
\label{fig:AxisDiagram1}
\end{figure}

In summary, the spacetime is incomplete both to the future and to the past, and extendible across both $v\pm \infty$. The extensions across $|v|\rightarrow \infty$ can be achieved in a standard fashion using a Kruskal-type of extension.
For the axis of symmetry with metric \eq{axis} we note that
$$
\gamma|_{x=0} =\frac{F(r)}{r^2+1} dv \left(dv+\frac{2(r^2+1)}{F(r)}dr \right)
$$
so that we can define a new coordinate $u$ by means of
\be
du := dv+\frac{2(r^2+1)}{F(r)} dr \label{u}
\ee
and the first fundamental form at the axis reads simply
\begin{equation}
\gamma|_{x=0} =\frac{F(r)}{r^2+1} du dv \,.
\label{1stff_axis}
\end{equation}
\begin{figure}[!ht]
\begin{center}
\includegraphics[height=6cm]{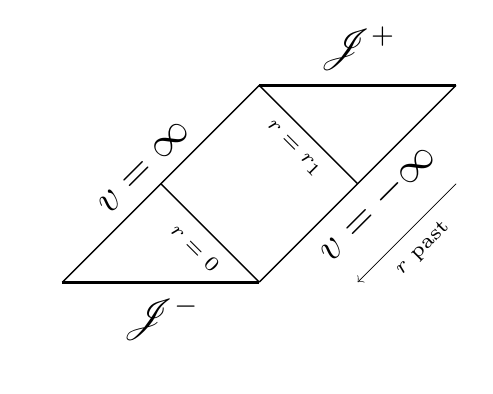}
\end{center}
\caption{\footnotesize{A conformal diagram of the axis $x=0$ similar to the previous one and with the same conventions, but now assuming that $\partial_r$ is past pointing (keeping $m>0$).  Again the intersection of the Killing horizons with the axis are the null geodesics labeled as $r=r_1$ and $r=0$, with $r_1<0$. Past null infinity $\scri^-$ is now defined by $r\rightarrow \infty$ while $\scri^+$ is given by $r\rightarrow -\infty$, and as before both $\scri$'s have the topology of $\mathbb{S}^3\setminus \{p\} =\mathbb{R}^3$. Null geodesics traveling along the vector field $-\partial_r$ extend all the way from $\scri^-$ to $\scri^+$ and are complete. The second family of null geodesics are incomplete to the future if they have $r>r_1$, while they are incomplete to the past if they have $r<0$ so that this Lorentzian manifold is extendible across the two borders marked by $v=\pm \infty$. It should be noted that this spacetime can be understood as a time reversal of that shown in figure \ref{fig:AxisDiagram1}. Moreover, one can smoothly join this diagram with that of Figure \ref{fig:AxisDiagram1} on {\em both} sides to produce an extension of the axis of the solution, as explained in the next Figure \ref{fig:AxisExtension}.}}
\label{fig:AxisDiagram2}
\end{figure}
Recalling \eq{F1}, 
the integral (\ref{u}) defining $u$ can be given in terms of elementary functions as (for $r_1\ne 0$, i.e.\ $m\ne 0$)
\begin{align}
& u-v  =\int \frac{2(r^2+1)}{F(r)} dr \nonumber
\\
&= \alpha_0 + \frac{3}{\Lambda 
r_1 (1 + 4r_1^2 + 3r_1^4) }\left [\frac{2 r_1^3(5+3 r_1^2)}{ \sqrt{4+ 3r_1^2}}\arctan\Big(\frac{2r + r_1}{\sqrt{4+ 3r_1^2}}\Big) + \right . \nonumber \\
& \left . 
2(1+r_1^2)^2\log|r-r_1|
-2(1+3 r_1^2)\log |r|  - r_1^2(r_1^2-1)\log(1+r^2 + r r_1 + r_1^2) \frac{}{}
\right ], \label{defr}
\end{align}
where $\alpha_0$ is an arbitrary integration constant.
We use  \eq{F1}  to eliminate either the  $\log|r-r_1|$- or the $\log|r|$-part, which yields 
\begin{align}
u-v 
= &
a_0 + c_1\arctan\Big(\frac{2r + r_1}{\sqrt{4+ 3r_1^2}}\Big) 
+ c_2\log |F(r)| 
+c_3\log |r-r_1| \nonumber  \\
& +c_4\log(1+r^2 + r r_1 + r_1^2)
\,,
\label{u-v1}
\end{align}
and
\begin{align}
u-v 
= &
 \hat a_0 + \hat  c_1\arctan\Big(\frac{2r + r_1}{\sqrt{4+ 3r_1^2}}\Big)
+\hat c_2\log  |F(r)|  +\hat c_3\log |r| \nonumber  \\
& 
+\hat c_4 \log(1+r^2 + r r_1 + r_1^2)
\,,
\label{u-v2}
\end{align}
respectively, for appropriately defined constants $c_i$ and $\hat c_i$ 
depending only on $r_1$ and $a_0$, $\hat{a}_0$ arbitrary constants.
Assuming $r_1< 0$, i.e.\ $m> 0$, (the other case is similar)
the constants $c_2$ and $\hat{c}_2$ satisfy 
\begin{equation*}
  c_2 >0, \quad \quad \hat{c}_2 <0.
\end{equation*}
%
We solve \eq{u-v1} and \eq{u-v2}  for $F(r)$,
\begin{align*}
 F(r)  
&=
- c_2^{-2}  e^{\frac{u-v}{c_2}}
|r-r_1|^{d_1}\underbrace{ D  \, 
|1+r^2 + r r_1 + r_1^2|^{d_2}
e^{d_3\arctan\big(\frac{2r + r_1}{\sqrt{4+ 3r_1^2}}\big) }}_{=:\mathfrak{F}(r)}
\\
 F(r)  
&=
- \hat{c}_2^{-2} e^{\frac{u-v}{\hat{c}_2}}
|r|^{\hat d_1}
\underbrace{\widehat D \,  |1+r^2 + r r_1 + r_1^2|^{\hat d_2}
e^{\hat d_3\arctan\big(\frac{2r + r_1}{\sqrt{4+ 3r_1^2}}\big) }
}_{=:\hat{\mathfrak{F}}(r)}
\end{align*}
for appropriately defined $d_i=d_i(r_1)$ and  $\hat d_i=\hat d_i(r_1)$,
and arbitrary non-zero constants $D$ and $\widehat{D}$.
Next, we insert both expressions  into \eq{1stff_axis},
\begin{eqnarray*}
\gamma|_{x=0} &=&\frac{-\mathfrak{F}(r) e^{\frac{u-v}{c_2}}|r-r_1|^{d_1}}{c_2^{2} (r^2+1)} du dv
\,,
\\
\gamma|_{x=0} &=&\frac{- \hat{\mathfrak{F}}(r) e^{\frac{u-v}{\hat{c}_2}}|r|^{\hat d_1}}{
\hat{c}_2^{2} (r^2+1)} du dv
\,.
\end{eqnarray*}
and introduce new coordinates
\begin{equation*}
U:=e^{\frac{u}{c_2}}\,, \quad V:= e^{-\frac{v}{c_2}}
\,, \quad \hat U:=e^{\frac{u}{\hat c_2}} \,, \quad \hat V:=  e^{-\frac{v}{\hat c_2}}
\end{equation*}
to end up with a Kruskal-type of extension,
\begin{equation}
\gamma|_{x=0} =\frac{\mathfrak{ F}(r) |r-r_1|^{d_1}}{r^2+1} dU dV
\,, \quad
\gamma|_{x=0} =\frac{\hat{\mathfrak{ F}}(r)|r|^{\hat d_1}}{r^2+1} dU dV
\,,
\end{equation}
where $\mathfrak{ F}(r)$ and $\hat{\mathfrak{ F}}(r)$ are non-singular in $r$.
Moreover, it is immediate to check that 
$r(UV)$ (resp. $r(\hat{U} \hat{V})$) defined by (\ref{defr}) is smooth near 
zero and satisfies $r(0) = 0$ (resp.  $r(0)=r_1$).
Thus, the first expression provides an extension through $v\rightarrow \infty$, 
at $r=0$
(note that this extended metric is degenerate at $r=r_1$), while the second one yields an extension through $v\rightarrow -\infty$, at $r=r_1$
(which is degenerate at $r=0$).

\begin{figure}[!ht]
\begin{center}
\includegraphics[height=6cm]{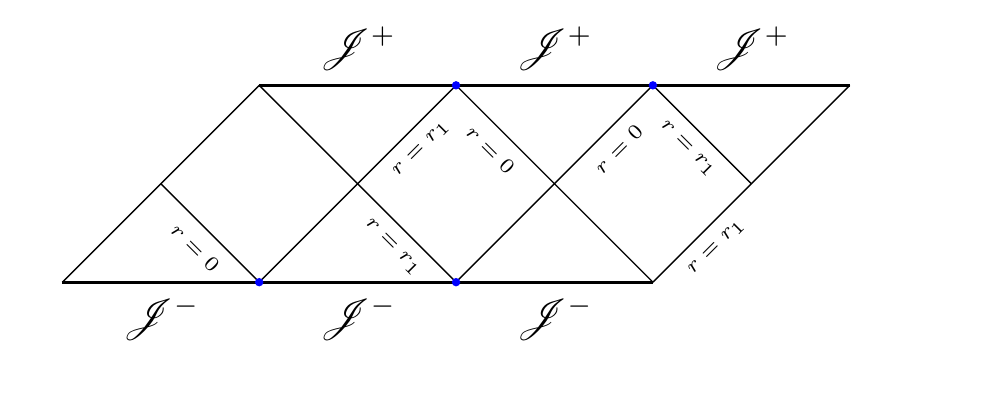}
\end{center}
\caption{\footnotesize{A conformal diagram of a maximal extension of the axis $x=0$ keeping the same conventions as before, but now with $\partial_r$ changing causal orientation depending on the patch ---and always keeping $m>0$. This extension is obtained by simply gluing a diagram of the type shown in \ref{fig:AxisDiagram2} to the left and right borders of the original diagram given Figure \ref{fig:AxisDiagram1}. Still, this extended axis can be further extended to the left and to the right, by now joining two copies, one at each side, of the diagram in Figure \ref{fig:AxisDiagram1}, and this process can be repeated indefinitely. The blue dots are not part of the $\scri$'s.}}
\label{fig:AxisExtension}
\end{figure}

 By using different patches of this type the metric at the axis can be extended indefinitely. This actually amounts to glue a metric of type \eq{axis} with a similar one with the sign of $r$ changed (this is shown in Fig.\ref{fig:AxisDiagram2}) across their boundaries $|v|\rightarrow \infty$. This process can be continued indefinitely and thus a maximal extension is then produced as shown in Figure \ref{fig:AxisExtension}.

\section{The metric at $\scri$}\label{Scri}
From the results in \cite{mpss2} we know that (i) 
$g$ admits a conformal completion {\em \`a la} Penrose with a conformally flat spacelike $\scri$, and (ii) that the topology of any of its connected components is $\mathbb{R}^3=\mathbb{S}^3\setminus \{p\}$. In fact, 
$\mathbb{S}^3\setminus \{p\}$ endowed with the standard round sphere 
in $\mathbb{S}^3$ is a representative of the conformal class 
of $\scri$.
In order to find an explicit representation of $\scri$
 it is enough to multiply $g$ by $1/r^2$ and then take the limit as $r\rightarrow \infty$. From the Kerr-Schild form of the metric shown in section \ref{Kerr-Schild} and the explicit form of the function $Q$ it is obvious that the metric $h$ at $\scri$ is
\bean
h :=\lim_{r\rightarrow \infty} \frac{1}{r^2} g = \lim_{r\rightarrow \infty} \frac{1}{r^2} \tilde g 
= \frac{3}{\Lambda} dx^2 + \frac{\Lambda}{3} \frac{1}{\cosh^2 x} \left(\mathrm{d}v^2+ \sinh^2 x \, d\Phi^2 \right).
\eean
By setting $y:= \sinh x = \tan \theta$, this adopts the explicitly conformally flat form
\begin{equation}
h = \frac{3}{\Lambda (1+y^2)} \left[ dy^2+\left(\frac{\Lambda}{3}\right)^2 \left(dv^2 + y^2 d\Phi^2 \right)\right].
\label{metric_at_scri}
\end{equation}
This metric is defined in all $\mathbb{R}^3$, which is diffeomorphic
to $\mathbb{S}^3 \setminus \{ p \}$ and we recover both the topology at
$\scri$ and the  statement concerning the conformal class.

Moreover, the results in \cite{mpss2} show that $\scri$ admits
a privileged conformal Killing vector $Y$ which determines
completely the TT tensor that, together with the conformal 
class, defines the asymptotic Cauchy data for the Einstein vacuum
field equations with positive $\Lambda$. This conformal Killing vector is
(the extension to $\scri$ of) the Killing vector
$\partial_t$ in the original form of the metric (\ref{KdS_limit}).
Implementing the coordinate changes we have introduced, it follows
that 
\begin{align*}
Y = \partial_v - \partial_{\Phi} 
\end{align*}
in the coordinates of (\ref{metric_at_scri}).  This conformal Killing vector
vanishes nowhere in $\mathbb{R}^3$. If we take the round sphere
representative  of the conformal class, then $Y$ vanishes at precisely 
one point  of $\mathbb{S}^3$ and this is the point $p$ that must be excluded
from $\scri$. It follows that $\mathbb{R}^3$ with the metric $h$
(\ref{metric_at_scri}) covers completely each connected component of
$\scri$. 

The fact that the topology at $\scri$ is
$\mathbb{S}^3 \setminus \{ p\}$
seems to be contradicted by  Fig. \ref{fig:AxisExtension} where 
two points are removed at the boundary of each connected component of $\scri$.
To  clarify this point, one must first realize that this figure corresponds 
only to the axis of symmetry, so  $\scri$ in the diagram
corresponds
to those points at infinity lying at the axis of symmetry of 
$\eta = \partial_{\Phi}$. This corresponds to the line $y=0$
in the metric $h$. The coordinate $v$ takes values in $(-\infty,\infty)$,
which is conformally diffeomorphic to an open interval $(-a,a)$.
The boundary of this interval has two points, and these are precisely 
the points depicted in the figure. The
single point $p$ in the spherical representation $\mathbb{S}^3
\setminus \{ p \}$ of $\scri$ can be transformed to many other objects
by changing the representative in the conformal class. In the representation
given by $h$ we may compactify the $v$ coordinate (e.g by
$V = \tanh v$), and $\scri$ is then 
a slab between two planes $V = -1$ and $V=1$. In this representation,
the boundary of $\scri$ consists of two parallel planes, as opposed
to the single
point $p$ in the spherical representation of the conformal class.
The intersection
of the slab with the axis of symmetry gives an open interval (hence with
two boundary points) and this is the representation that is depicted 
in Fig \ref{fig:AxisExtension}.



\section{Conclusions}

The $a \rightarrow \infty$-KdS-limit spacetime analyzed in this paper
has a number of interesting properties, both by itself and in relation to
the limit process from the Kerr-de Sitter family. In particular,
the limit $a \rightarrow \infty$
changes some properties of the spacetime in a discontinuous way.
As discussed in the text, the cross sections of the Killing 
horizons have the topology of a disk  in the 
$a \rightarrow \infty$-KdS-limit spacetime, while they are topologically $\mathbb{S}^2$
for the Kerr-de Sitter spacetime. We have also found that these disks are unbounded
but have finite total area, which happens to be half the limit of the areas 
of the $\mathbb{S}^2$ sections of Kerr-de Sitter. 
The same drop by a factor of $2$ is found in the total angular momentum of the Killing horizon
not touching the singularity. We may understand all this as follows.

The surfaces with $v$ and $r$ constants, which are compact for finite $a$, 
are deformed into more and more oblate shape, with larger and larger equatorial planes, and then in the limit when $a \rightarrow \infty$
they just ``break" and become infinite disks (the equatorial circle having been pushed to infinity). Among this set of surfaces, those on the horizons are the only ones keeping a finite area. Still, this ``breaking'' is the underlying reason
why the finite area of these surfaces on the horizons in the limit spacetime drops by a factor of two. This also explains the
drop in the total angular momentum of the horizon that
stays away from the singularity.  The fact that
the shapes get more and more oblate is physically consistent with the fact that
we are pushing the specific angular momentum $a$ to very large values. 
During the limiting process, the singularity at $r=0$ and 
$\theta =\frac{\pi}{2}$, which is timelike for Kerr-de Sitter,
gets closer and closer to one of the horizons, and 
at the same time gets more and more tilted, until it actually ends up, in the limit, at the horizon $r=0$ and becomes a null-like surface. 

\section*{Acknowledgments}
We thank J. Podolsk\'y for his valuable comments and useful suggestions.
MM acknowledges financial support under the projects
FIS2015-65140-P (Spanish MINECO/FEDER) and
SA083P17 (JCyL)
TTP acknowledges financial support by the Austrian Science Fund (FWF)
P~28495-N27.
J.M.M.S. is supported under Grants No. FIS2014-57956-P (Spanish MINECO-fondos FEDER), No. IT956-16 (Basque Government), and EU COST action No. CA15117 ``CANTATA.'' 

\section*{Appendix}
We have learnt, for example due to the structure of $\scri$, that the metric $\tilde g$ in \eq{deSitter_EF} can cover only a portion of the de Sitter spacetime. In this appendix we provide the necessary changes of coordinates taking
\eq{deSitter_EF} to the global de Sitter spacetime. 
As an intermediate step set
%
\begin{equation}
\hat \mu =\sqrt{\frac{\Lambda}{3}}\, v-\sqrt{\frac{3}{\Lambda}}\,\frac{1}{r} \,, \quad \hat\varphi =\frac{\Lambda}{3} \Phi -\arctan r
\,,
\end{equation}
so that \eq{deSitter_EF} becomes
\begin{eqnarray}
\tilde g  &= & -\frac{3}{\Lambda}\frac{\mathrm{d} r^2}{r^2}
 +r^2\cos^2\theta\mathrm{d}\hat \mu^2 
+\frac{3}{\Lambda} \cos^2\theta\frac{\tan^2\theta}{r^2(1+r^2)}\mathrm{d}r^2
\nonumber
\\
&&
+\frac{3}{\Lambda}(1 +r^2) \sin^2\theta\mathrm{d}\hat\varphi^2
+\frac{3}{\Lambda}(r^2+ \cos^2\theta)\frac{\mathrm{d}\theta^2}{\cos^2\theta}
\,,
\end{eqnarray}
and define now
\begin{equation}
\hat t = \sqrt{\frac{3}{\Lambda}}\,\log (r\cos\theta)\,, \quad \hat \rho = \sqrt{\frac{3}{\Lambda}}\,\frac{\sqrt{1+r^2}}{r}\tan\theta
\,,
\end{equation}
leading to the well-known ``steady-state'' coordinate system which covers half of the de Sitter spacetime,
\begin{eqnarray}
g_{\mathrm{dS}/2}  &= & -\mathrm{d}\hat t^2 + e^{2\sqrt{\frac{\Lambda}{3}}\, \hat t}\Big(\mathrm{d}\hat \mu^2 + \mathrm{d}\hat\rho^2 + \hat\rho^2\mathrm{d}\hat\varphi^2\Big)
\,.
\label{deSitter_half}
\end{eqnarray}
The extension transformation into global coordinates is (cf.\ e.g.\ \cite{griffiths}),
\begin{eqnarray}
\hat t &=& \sqrt{\frac{3}{\Lambda}}\,\log\Big(\sinh\big(\sqrt{\frac{\Lambda}{3}}\, T\big) + \cosh\big(\sqrt{\frac{\Lambda}{3}}\, T\big) \cos\chi\Big)
\,,
\\
\hat\mu &=& 
\frac{  \sqrt{\frac{3}{\Lambda}}\,\sin\chi\cos\Theta}{\tanh\big(\sqrt{\frac{\Lambda}{3}}\, T\big) +  \cos\chi}
\,,
\\
\hat\rho &=&  \frac{
 \sqrt{\frac{3}{\Lambda}}\, \sin\chi\sin\Theta
}{\tanh\big(\sqrt{\frac{\Lambda}{3}}\, T\big) + \cos\chi}
\,,
\\
\hat\varphi &=& \tilde\Phi
\,.
\end{eqnarray}
Altogether, the following extension 
%
%
\begin{eqnarray}
 T &=&
\sqrt{\frac{3}{\Lambda}}\mathrm{arsinh}\Big[ \frac{\Lambda}{6} v\Big(\frac{\Lambda}{3}vr - 2\Big)  \cos\theta
+\frac{r}{2\cos\theta} \Big]
\,,
\label{coord_trafo_1}
\\
\chi &=&\arccos\left[ \frac{r\cos^2\theta -\frac{r}{2}  -  \frac{\Lambda}{6} v\Big(\frac{\Lambda}{3}vr - 2\Big)  \cos^2\theta
}{ \sqrt{\cos^2\theta +\Big(\frac{r}{2} + \frac{\Lambda}{6} v\big(\frac{\Lambda}{3}vr - 2\big)  \cos^2\theta \Big)^2} } 
\right]
\,,
\\
\Theta &=& \arctan\Big( \frac{ \sqrt{1+r^2}}{\frac{\Lambda}{3} vr- 1}\tan\theta\Big) 
\,,
\\
\tilde\Phi &=& \frac{\Lambda}{3} \Phi -\arctan r
\,.
\label{coord_trafo_4}
\end{eqnarray}
accomplishes the desired transformation from  \eq{deSitter_EF} the global de Sitter line element 
\begin{equation}
g_{\mathrm{dS}} = -\mathrm{d}T^2 +\frac{3}{\Lambda}\cosh^2\big( \sqrt{\frac{\Lambda}{3}}\,T\big) \Big(\mathrm{d}\chi^2 + \sin^2\chi(\mathrm{d}\Theta^2 + \sin^2\Theta\mathrm{d} \tilde\Phi^2)\Big)
\,.
\label{deSitter_global}
\end{equation}
%

%



\begin{thebibliography}{[10]} 
\bibitem{Ashtekhar}
A. Ashtekar, C. Beetle, J. Lewandowsky: \emph{Geometry of generic
isolated horizons}, Class. Quantum Grav. {\textbf 21} (2002) 2549-2740.
\bibitem{Cho}
J.-H. Cho, Y. Ko, S. Nam: \emph{The entropy function
for the extremal Kerr-(anti-)de Sitter Black Holes},
Annals Phys. {\bf  325} (2010) 1517-1536.
\bibitem{Chrusciel} P.T. Chru\'sciel, L. Ifsits: \emph{The cosmological constant and the energy of gravitational radiation},  Physical Review \textbf{D 93} (2016) 124075.
\bibitem{griffiths} J.B. Griffiths, J. Podolsk\'y: \emph{Exact spacetimes in Einstein's general relativity}, Cambridge: Cambridge University Press, 2009.
\bibitem{Jaramillo} 
E. Gourgoulhon, J.L. Jaramillo: 
\emph{A 3+1 perspective on null hypersurfaces and isolated horizons},
Phys. Rept. {\bf 423} (2006) 159-294.
\bibitem{MarsTotally} M. Mars: \emph{Stability of MOTS in totally geodesic null
horizons}, Class Quantum Grav. {\textbf 29} (2012) 145019.
\bibitem{mpss} M. Mars, T.-T. Paetz, J.M.M. Senovilla, W. Simon: \emph{Characterization of (asymptotically) Kerr-de Sitter-like spacetimes at null infinity}, 
Class. Quantum Grav. \textbf{33} (2016) 155001.
\bibitem{mpss2} M. Mars, T.-T. Paetz, J.M.M. Senovilla: \emph{Classification of Kerr-de Sitter-like spacetimes with conformally flat $\scri$}, Class. Quantum Grav. {\bf 34} (2017) 095010
\bibitem{MS} M. Mars and J.M.M. Senovilla, \emph{Axial symmetry and conformal Killing vectors}, Class. Quantum Grav. {\bf 10} (1993) 1633--1647
\bibitem{mars_senovilla} M. Mars, J.M.M. Senovilla: \textit{A spacetime characterization of the Kerr-NUT-(A)de Sitter and related metrics}, Ann. Henri Poincar\'e {\bf 16} (2015) 1509--1550.
\bibitem{oelz} C. R. \"Olz: \emph{The global structure of Kerr-de Sitter
metrics}, diploma thesis (2013),
\url{http://othes.univie.ac.at/29183/1/2013-03-28_0607354.pdf}.
\bibitem{S2002} J.M.M. Senovilla, \emph{Trapped surfaces, horizons and exact solutions in higher dimensions}, Class. Quantum Grav. {\bf 19} (2002) L113
\bibitem{Szabados} L.B. Szabados, P. Tod: \emph{
A positive Bondi-type mass in asymptotically de Sitter spacetimes},
Class. Quantum Grav. {\bf 32} (2015) 205011.

\end{thebibliography}
\end{document}